\def\spose#1{\hbox to 0pt{#1\hss}}
\def\approxlt{\mathrel{\spose{\lower 3pt\hbox{$\sim$}}
	\raise 2.0pt\hbox{$$<$$}}}
\def\approxgt{\mathrel{\spose{\lower 3pt\hbox{$\sim$}}
	\raise 2.0pt\hbox{$>$}}}

\def\multleft#1{\hbox to size{\vbox {\halign {\lft{##}\cr #1}}\hfill}\par}
\def\multright#1{\hbox to size{\vbox {\halign {\rt{##}\cr #1}}\hfill}\par}

\def\today{\ifcase\month\or January\or February\or March\or April\or May\or
      June\or July\or August\or September\or October\or November\or December\fi
      \space\number\day, \number\year}
\def\$<${\thinspace}
\def\s{\hbox{\phantom{5}}}	

\def\boxit#1{\vbox{\hrule\hbox{\vrule\kern3pt\vbox{\kern3pt
          #1 \kern3pt}\kern3pt\vrule}\hrule}}

\def\cm{{\rm\thinspace cm}}

\def\erg{{\rm\thinspace erg}}

\def\keV{{\rm\thinspace keV}}
\def\km{{\rm\thinspace km}}

\def\Mpc{{\rm\thinspace Mpc}}
\def\Msun{\hbox{$\rm\thinspace M_{\odot}$}}

\def\s{{\rm\thinspace s}}
\def\yr{{\rm\thinspace yr}}
\def\sr{{\rm\thinspace sr}}

\def\ergpcmsqps{\hbox{$\erg\cm^{-2}\s^{-1}\,$}}

\def\ergps{\hbox{$\erg\s^{-1}\,$}}

\def\kmps{\hbox{$\km\s^{-1}\,$}}

\def\psqcm{\hbox{$\cm^{-2}\,$}}

\def\kmpspMpc{\hbox{$\kmps\Mpc^{-1}$}}

\documentstyle[psfig]{mn}
\begin{document}
\hsize=6truein

\title{A physical model for the hard X-ray background}

\author[]
{\parbox[]{6.in} {R.J.~Wilman$^1$, A.C.~Fabian$^1$ and P.E.J.~Nulsen$^2$ \\
\footnotesize
1. Institute of Astronomy, Madingley Road, Cambridge CB3 0HA \\ 
2. University of Wollongong, Wollongong NSW 2522, Australia \\ }}
\maketitle

\begin{abstract}
We combine a semi-analytic galaxy formation model (Nulsen \& Fabian~1997) with a prescription for the obscured growth of massive black holes (Nulsen \& Fabian~2000; Fabian~1999), to reproduce the hard X-ray background (XRB), the local 2--10\keV~AGN luminosity function and the source counts, including recent {\em Chandra} results. The model also complies with constraints on the AGN contribution to the far-infrared and sub-mm backgrounds.

The co-moving density of luminous AGN ($L(2-10\keV) > 10^{44}$\ergps, in the unabsorbed rest-frame) in the model declines sharply since $z=2$, mimicking the observed evolution of the quasar population. The abundance of lower luminosity AGN simultaneously increases, and we identify such objects with the Seyfert galaxies. These features reflect an epoch-dependent rate of Bondi accretion from the hot, cooling atmosphere: at $z<2.5$ most AGN are fed at much less than 10 per cent of the Eddington rate, whilst at $z>2.5$ most accrete at 10--50 per cent Eddington. Unlike other models for quasar and galaxy formation, we do not consider major mergers between normal galaxies.

The model produces an excess soft XRB below 4\keV~which can be removed by modifying the wind expulsion model of Fabian~(1999) to incorporate anisotropic ejection.
\end{abstract}


\section{INTRODUCTION}
Since the discovery of the cosmic X-ray background (XRB) by Giacconi et al.~(1962), almost every class of astronomical X-ray source has at some stage been invoked to account for it. The present consensus is that it represents the emission from active galactic nuclei (AGN) integrated over cosmic time; in the soft 0.5--2.0\keV~band, deep {\em ROSAT} surveys have established that the bulk of it is due to broad line AGN i.e. quasars and Seyfert 1 galaxies (Hasinger et al.~1998; Schmidt et al.~1998), but the situation in the harder 2--10\keV~band is quite different. No more than 30 per cent of it was resolved by the {\em ASCA} and {\em BeppoSAX} satellites, with the optical identifications being a mixture of quasars and narrow emission-line galaxies (Boyle et al.~1998, Ueda et al.~1999 and Akiyama et al.~2000; Fiore et al.~1999). Since the realisation by Setti \& Woltjer~(1989) that the flat spectrum of the 2--10\keV~XRB implies considerable absorption in most objects, many workers have constructed AGN synthesis models to account for the remainder of the hard XRB (see e.g. Madau, Ghisellini \& Fabian~1994; Comastri et al.~1995; Wilman \& Fabian~1999). Fabian \& Iwasawa~(1999) emphasised the importance of the XRB as a probe of the cosmological evolution of the black hole population, by demonstrating that the XRB sources plausibly account for all of the accretion power in the Universe and that most of it is intrinsically absorbed.

With the advent of the {\em Chandra} and {\em XMM} telescopes, the predictions of these models for the 2--10\keV~source counts below fluxes of $10^{-14}$\ergpcmsqps~are now being tested. Mushotzky et al.~(2000) recently reported the results of an extremely deep {\em Chandra} observation with which they determined the source counts down to $4 \times 10^{-15}$\ergpcmsqps~and thereby resolved essentially all of the XRB in this band (see also Brandt et al.~2000). But their optical follow-up campaign identified two new classes of object: (1) optically `faint' objects ($I \gg 23$ mag) which may be either the first quasars at very high redshifts or dust enshrouded AGN at redshifts $z>2$; (2) point-like hard X-ray sources in the nuclei of apparently `normal' bright galaxies showing no other signs of activity. In a related vein, Fabian et al.~(2000) cross-correlated {\em Chandra} fields containing faint serendipitous sources with SCUBA sub-millimetre observations of the same lensing cluster fields; only one source is common to both datasets, suggesting that if the SCUBA sources do host powerful AGN, they must either contribute a small fraction of the sub-mm power, or be Compton-thick with X-ray scattering fractions of less than 1 per cent. This result was confirmed by Hornschemeier et al.~(2000), who detected none of the 10 sub-mm sources in and around the Hubble Deep Field North in a 166-ks {\em Chandra} exposure. We stress, however, that {\em Chandra} has so far resolved the bulk of the XRB in the 2--7\keV~band, which is mainly due to Compton-thin sources; forthcoming deep surveys with {\em XMM} will shed light on the more heavily obscured sources which are likely to contribute up to 15\keV, but the bulk of the energy density of the XRB at $\sim 30$\keV -- which is likely to be dominated by Compton-thick sources-- will not be resolved until {\em Constellation-X} is in flight (see e.g. Valinia et al.~1999).

The findings of Mushotzky et al.~(2000) and Fabian et al.~(2000) demonstrate that the newly-discovered XRB sources may be quite different from established classes of AGN, such as the quasars and, at low redshift, the Seyfert galaxies. Indeed, Fabian~(1999) has argued that the new XRB sources represent the hitherto unobserved phase associated with the major growth of massive black holes, whereas classical (optically-selected) quasars and Seyfert galaxies represent a later, transient phase during which the black hole acquires little additional mass. In his model for the formation of galactic bulges and central black holes, young spheroidal galaxies have a significant distributed component of cold dusty clouds which leads to absorption of X-rays from the accreting black hole. The accretion is terminated by wind-driven gas expulsion when the mass of the black hole reaches a critical fraction of that in the surrounding spheroid (see also Silk \& Rees~1998), thereby accounting for the observed correlation between the mass of the remnant black hole and its host spheroid (Magorrian et al.~1998; Van der Marel 1999). Following expulsion, the object shines briefly as an optically-selected quasar. 

In this paper, we incorporate Fabian's (1999) model for the obscured growth of massive black holes into the semi-analytic model for galaxy formation and quasar fuelling developed by Nulsen \& Fabian~(1997, 2000) (hereafter NF97 and NF00). The result is a physical model for the XRB which works forward in cosmic time from the epoch when the first massive bound objects were forming, in contrast to the aforementioned synthesis models which start with the locally observed AGN luminosity functions and extrapolate them back to earlier epochs. We assume that quasars are fuelled by Bondi accretion from the dense, hot cooling atmospheres which form around collapsing objects about the size of normal galaxies. Earlier attempts to incorporate AGN into the framework of semi-analytic models were made by Kauffmann \& Haehnelt~(2000) and Cattaneo~(2000), but they assumed that black holes are fed by cold gas during major galaxy mergers and they did not use the XRB as a constraint.

\section{DESCRIPTION OF THE MODEL}
We use the semi-analytic galaxy formation model and associated scheme for quasar fuelling developed by NF97 and NF00, the salient features of which we outline here. 

\subsection{Galaxy formation}
The Cole \& Kaiser~(1988) block model is used to simulate the hierarchical growth of clustering, while the behaviour of the gas in collapsed regions is simulated using the model of Nulsen \& Fabian~(1995). In short, the gas in the collapsed halo is split into two regions, according to the value of $\tau=t_{\rm{cool}}/t_{\rm{grav}}$, the ratio of the cooling time of the gas to its free-fall time to the centre of the halo. Within a radius $R=R_{\rm{CF}}$ where $\tau<1$, the gas cools rapidly and forms stars. Any gas at $R>R_{\rm{CF}}$ has $\tau>1$ and is assumed to participate in a cooling flow (Fabian 1994). Collapses with and without the cooling flow atmosphere are identified as normal galaxies and dwarf galaxies, respectively.

The burst of star formation in $R<R_{\rm{CF}}$ leads quickly to supernovae, which can expel some or all of the remaining gas from the system; some gas is always expelled from dwarf galaxies, and in normal protogalaxies hot gas which is not ejected is heated and enriched by the supernovae. A normal galaxy is a spiral galaxy if the hot gas cools completely before the present or the next hierarchical collapse, otherwise it is an elliptical galaxy. Any collapse with at most one infalling normal galaxy forms a normal galaxy, so that dwarf galaxies are assumed to be destroyed in a collapse, with their stars contributing to the spheroid of the resulting system. A system containing more than one normal galaxy is assumed to form a group or cluster, as mergers between normal galaxies are ignored. Any collapse that is followed less than one dynamical time later by further collapse is ignored.

We use an open CDM cosmology, with $H_{\rm{0}}=50$\kmpspMpc, density parameter $\Omega=0.3$, baryon density parameter $\Omega_{\rm{b}}=0.075$ and $\sigma_{8}=1$.

\subsection{Quasars: fuelling, obscuration and demise}
NF00 proposed that quasars are formed and fed largely by the accretion of hot gas from the atmospheres which form around normal galaxies. They showed that if such an atmosphere forms a nearly ``maximal'' cooling flow, then a central black hole can accrete at close to its Eddington limit, leading to the exponential growth of a seed black hole. They also incorporated this model into the above semi-analytic galaxy formation model, which we adopt here with some modifications.

Each of the smallest block model units has a mass of $1.5\times 10^{10}$\Msun, and contains at its centre a seed black hole of mass $M_{\rm{h(seed)}}$. When a block collapses, the black holes associated with all merging sub-blocks are assumed to merge into a single black hole. The resulting nuclear black hole grows exponentially by Bondi accretion from the cooling flow, on the timescale given by eqn.~(8) of NF00. Whilst accreting in this manner, the black hole is assumed to radiate with an accretion efficiency of 10 per cent (i.e. $L_{\rm{Bol}}=0.1\dot{M_{h}} c^{2}$, $\dot{M_{h}}$ being the accretion rate), with an intrinsic 2--10\keV~luminosity equal to 3 per cent of the bolometric one (as deduced from the work of Elvis et al.~1994). The intrinsic spectrum is taken to be a power-law of photon index $\Gamma=1.9$, with an exponential cut-off at $E_{\rm{c}}=360\keV$; for objects with $L(2-10\keV)<10^{44}$\ergps, we added to this a component reflected from a cold accretion disc subtending $2\pi\sr$ at the source, for a fixed inclination angle of $60^{\circ}$. More powerful objects (i.e. quasars, rather than Seyfert galaxies) are not observed to exhibit such reflection components, perhaps because the accretion disc is ionized (see Reeves \& Turner~2000 and references therein). 

The above intrinsic spectrum is then absorbed by the cold, dusty clouds deposited by the cooling flow, as suggested by Fabian~(1999). He showed that for a density distribution of cooled gas $\rho \propto r^{-2}$, the column density presented by such material exterior to a radius $r_{\rm{in}}$ is given by:
\begin{equation}
N_{\rm{H}} = \frac{v^{2}f}{2 \pi G m_{\rm{p}} r_{\rm{in}}}
\end{equation}
where $f$ is the mass fraction in cold clouds and $v$ the line of sight velocity dispersion of the isothermal spheroid. For $r_{\rm{in}}$ we take the radius $r_{\rm{x}}$ at which the Bondi solution takes over from the cooling flow, given by NF00 as:
\begin{equation}
r_{\rm{x}} = \frac{G M_{\rm{h}}}{2 s_{\rm{i}}^{2} {\cal M}_{\rm{i}}^{1/2}} 
\end{equation}
where $M_{\rm{h}}$ is the mass of the black hole, ${\cal M}_{\rm{i}}$ the Mach number of the cooling flow and $s_{\rm{i}}$ the adiabatic sound speed at large $r$. Within $r_{\rm{x}}$ the flow comes under the influence of the black hole and the flow time (in to the black hole) becomes negligible compared with the cooling time of the gas, so there is effectively no further cooling. Using eqns.~(1) and (2), $s_{\rm{i}}^{2}=2 \gamma v^{2}$ for an isothermal cooling flow (hereafter CF) (with $\gamma=5/3$), the normalization adopted by Fabian~(1999) for the Faber-Jackson relation between $v$ and the mass of the host spheroid $M_{\rm{sph}}$ ($\propto v^{4}$), and a hydrogen mass fraction of $X=0.75$, we find the following expression linking the obscuring column density to the instantaneous mass of the accreting black hole, $M_{\rm{h}}$:
\begin{equation}
N_{\rm{H}} = 0.042 f {\cal M}_{\rm{i}}^{1/2} \frac{M_{\rm{sph}}} {M_{\rm{h}}} N_{\rm{T}}
\end{equation}
where the time dependence of ${\cal M}_{\rm{i}}$ may be taken from eqn.~(12) of NF00 and $1/N_{\rm{T}}=8.07 \times 10^{-25}$cm$^{2}$ is the Thomson electron scattering cross-section per hydrogen atom. The obscuration therefore diminishes as the black hole grows. We make use of the Monte Carlo simulations of Wilman \& Fabian~(1999) to compute the X-ray spectrum transmitted through a sphere of absorbing material of thickness $N_{\rm{H}}$. Both photoelectric absorption from neutral material and Compton down-scattering are modelled, the latter being described by the Klein-Nishina cross-section. We use an iron abundance of 5 times the solar value, after Wilman \& Fabian~(1999) demonstrated that this leads to better fits to the XRB spectrum and creates additional parameter space. The value of $M_{\rm{sph}}$ used in eqn.~(3) is fixed prior to the onset of the accretion -- i.e. we neglect any contribution from baryonic dark matter deposited by the cooling flow during the accretion.

There are three ways in which the growth of the black hole can be arrested: (i) exhaustion of the hot gas supply; (ii) participation in a new collapse; (iii) the black hole mass could reach a critical fraction of that of the surrounding spheroid, such that the wind which it drives is able to expel the remaining gas, as proposed by Fabian~(1999). He showed that for reasonable wind parameters, this can be made to happen when $M_{\rm{h}}=M_{\rm{h}(crit)}=0.005 M_{\rm{sph}}$, in agreement with the findings of Magorrian et al.~(1998) from the demographics of remnant supermassive black holes in nearby galaxies. Moreover, the model predicts that $M_{\rm{h}} \propto v^{4}$, in accord with the recent findings of Gebhardt et al.~(2000). Following this expulsion, the black hole is assumed to shine as an unobscured, optically-selected, quasar for a time $\Delta t_{\rm{qso}}$ (or until the next collapse if that happens sooner), with a constant luminosity equal to that immediately prior to expulsion. Unlike NF00, we do not assume that objects make a transition to an advection dominated (ADAF) phase when the accretion rate drops below a critical value.

In the next section we investigate the consequences of the model for the X-ray background and the quasar population, but before doing so we point out its deficiencies. Firstly, the Cole \& Kaiser~(1988) block has the drawback that halo masses always grow in discrete steps by factors of 2. As Somerville \& Kolatt~(1999) mention, this is problematic for semi-analytic galaxy formation modelling where one would like to follow individual systems with higher resolution; they propose a merger tree formalism which overcomes this and other difficulties, but for simplicity we do not use it here. There are also complications in the fuelling and obscuration model, related to the effect of the cold clouds on the structure of the CF. These depend upon whether the cold clouds rapidly decouple from the flow (as assumed here and in NF00) or remain tied to it, by e.g. magnetic fields (see e.g. Nulsen~1986). In the latter case, the cold clouds increase the total gas density but it is the {\em hot phase alone} which provides the pressure to support the flow against gravity and maintain approximate hydrostatic equilibrium; as a result, the accretion rate can be significantly higher than that given by eqn.~(A9) of NF00. Our parameterisation of the obscuration in terms of a constant $f$, the mass fraction in cold clouds, is also quite simplistic. In practice $f$ is likely to be a function of time, with the obscuration being determined by a competition between the amount of cold material deposited by the CF and the position of the black hole accretion radius ($\propto M_{\rm{h}}$; see eqn.~(2)). Such non-trivial refinements to the model will be addressed in a later paper, but for now it is sufficient to note that our current treatment is likely to be accurate at the most important time, during the last black hole doubling time when the quasar is at its most powerful and the obscuration smallest (i.e. immediately prior to expulsion).

\section{RESULTS}
To demonstrate that the model works satisfactorily, we present results for a run comprising 60 iterations of the block model with the following parameter values: $M_{\rm{h(seed)}}=1.6 \times 10^{6}$\Msun, $f=0.37$ and $\Delta t_{\rm{qso}}=9 \times 10^{7}$\yr. At present, we leave $M_{\rm{h}(crit)}/M_{\rm{sph}}$ at the Magorrian et al.~(1998) value of 0.005, despite more recent work suggesting that the true figure may be closer to 0.003 (see e.g. van der Marel~1999). With the exception of $\Delta t_{\rm{qso}}$ (see discussion in section~4), the numerical values of these parameters accord well with those suggested by Fabian~(1999).

\begin{figure}
\psfig{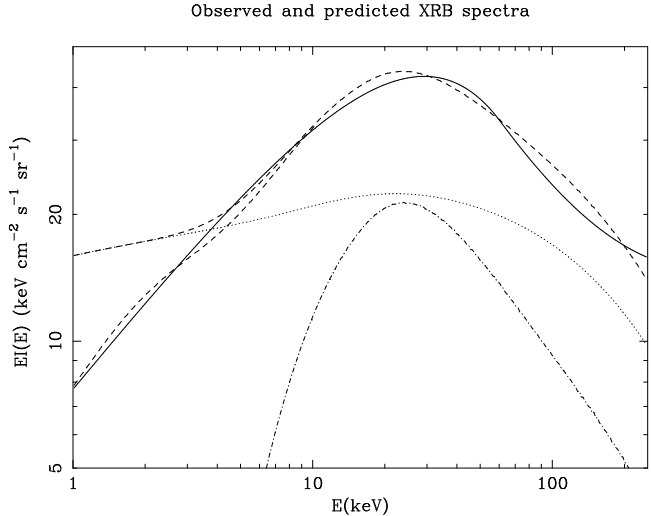}
\caption{\normalsize The upper dashed line shows the XRB spectrum generated by the model of section 3, with the contributions of obscured and unobscured active nuclei indicated by the dot-dashed and dotted lines, respectively. The solid line is the analytical fit of Gruber~(1992) to the observed XRB spectrum. The lower dashed line shows the effect of introducing $N_{\rm{H}}=10^{22.75}$\psqcm~of absorption into 70 per cent of the previously unobscured AGN spectra (for all luminosities).}
\label{fig:spec1}
\end{figure}

Fig.~\ref{fig:spec1} shows the XRB spectrum generated by this model. The soft excess below 4\keV~can easily be removed by the introduction of some moderate absorption ($N_{\rm{H}} \sim 10^{22-23}$\psqcm) into the spectra of some of the (currently unobscured) objects: e.g. the figure demonstrates that this can be effected by obscuring 70 per cent of the unobscured sources by $N_{\rm{H}}= 10^{22.75}$\psqcm~(for all luminosities). Wind expulsion is likely to be a gradual and/or anisotropic process, with the quasar passing through an intermediate phase between the extremes of being Compton-thick and totally unobscured, as discussed in section 4.

\begin{figure}
\psfig{figure=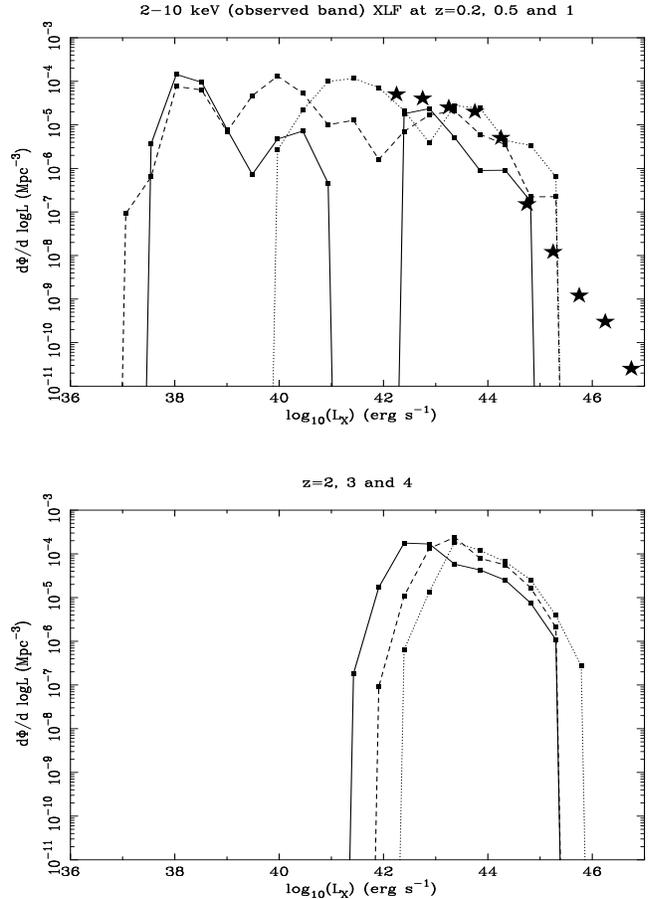,width=0.48\textwidth,angle=270}
\caption{\normalsize 2--10\keV~(in the observed frame) luminosity functions generated by the model of section 3, at redshifts 0.2 (solid line), 0.5 (dashed) and 1 (dotted) in the upper panel, and at redshifts 2 (solid), 3 (dashed) and 4 (dotted) in the lower panel. The filled stars in the upper panel represent the $z=0$ Type 1 + Type 2 AGN XLF derived by Boyle et al.~(1998) using a maximum-likelihood analysis of {\em ASCA} and {\em HEAO-A1} data (but for a $q_{\rm{0}}=0.5$ cosmology).}
\label{fig:xlf1}
\end{figure}

Fig.~\ref{fig:xlf1} shows the 2--10\keV~luminosity functions (XLFs) generated by the model. Those at $z \leq 1$ are compared with the $z=0$ AGN XLF deduced by Boyle et al.~(1998) from a combination of {\em ASCA} and {\em HEAO-A1} data (but for a $q_{0}=0.5$ cosmology). The double-peaked structure of the XLFs at $z \leq 1$ (which is most pronounced at $z=0.2$ where there are no objects at $L_{\rm{X}} \sim 10^{41.5}$\ergps) is due to the separate contributions of the unobscured and obscured objects at high and low luminosities, respectively. As discussed in section 4, at all epochs most of the obscured objects are Compton-thick with $N_{\rm{H}} \sim 3N_{\rm{T}}$, from which there is very little transmitted flux in the {\em observed} 2--10\keV~band for low redshift objects; at the higher redshifts in the lower panel of Fig.~\ref{fig:xlf1}, a favourable negative K-correction applies to the {\em observed} 2--10\keV~flux of a moderately Compton-thick source (as discussed by Wilman \& Fabian~1999), such that the contributions of the obscured and unobscured AGN overlap in luminosity. There is some scope for adjusting the form of the XLF at $z \leq 1$ by introducing some {\em scattered} flux into the spectra of the obscured objects (of which there is at present none).

\begin{figure}
\psfig{figure=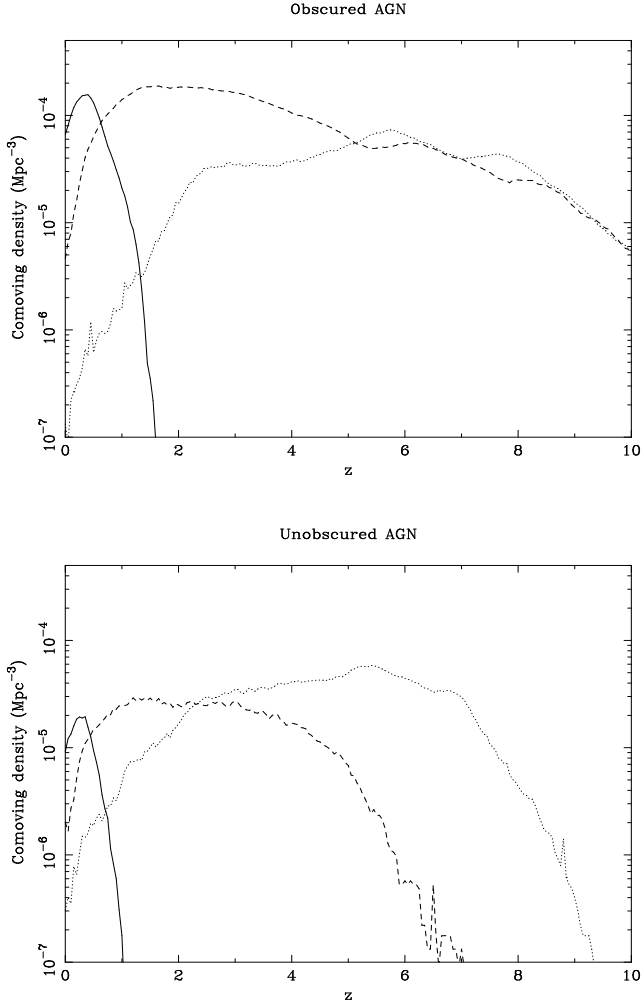,width=0.48\textwidth,angle=270}
\caption{\normalsize Redshift evolution of the co-moving densities of obscured and unobscured active nuclei, broken down according to the intrinsic 2--10\keV~luminosity ($L_{\rm{X}}$) (i.e. unabsorbed and in the rest-frame). The solid, dashed and dotted lines show objects with $L_{\rm{X}} \leq 10^{43}$, $10^{43} < L_{\rm{X}} \leq 10^{44}$ and $10^{44} < L_{\rm{X}} \leq 10^{45}$\ergps, respectively; in the lower panel, the spikes at $z \sim 6.5$ (dashed curve) and $z \sim 9$ (dotted curve) are due to numerical fluctuations.}
\label{fig:evol1}
\end{figure}

Our model fails, however, to produce enough of the high luminosity objects with $L_{\rm{X}}>10^{45}$\ergps. Such objects would not contribute significantly to the XRB, but we speculate that they could be fuelled by mergers between normal galaxies, which the model in its current form does not treat (see section~4). In addition, systems with two or more normal galaxies represent groups and clusters in which the intragroup or intracluster medium can increase the fuelling rate (especially if Compton feedback occurs; Fabian \& Crawford~1990). Concerning evolution, Boyle et al.~(1998) parameterized it with pure luminosity evolution of the form $(1+z)^{k}$ for $z \leq 2$, with $k \simeq 2$, but our model appears to show some density evolution out to $z=2$. Fig.~\ref{fig:evol1} demonstrates more readily how the population evolves, by showing the co-moving densities of the obscured and unobscured objects as functions of redshift, broken down according to the 2--10\keV~luminosity in the unabsorbed rest-frame. Two features are immediately apparent: the first is that the co-moving density of luminous ($L_{\rm{X}}>10^{44}$\ergps) unobscured AGN (which we identify with optically-selected quasars) increases by more than two orders of magnitude between the present epoch and $z=2$, mirroring the observed evolution of the quasar luminosity function (its observed behaviour beyond $z \simeq 3$ remains controversial: see Miyaji et al.~2000, Schmidt et al.~1995 and Shaver et al.~1999 for perspectives at X-ray, optical and radio wavelengths, respectively); secondly, we also see how low luminosity active nuclei become greatly more abundant towards the present epoch, and we tentatively identify these objects with the Seyfert galaxies.

\begin{figure}
\psfig{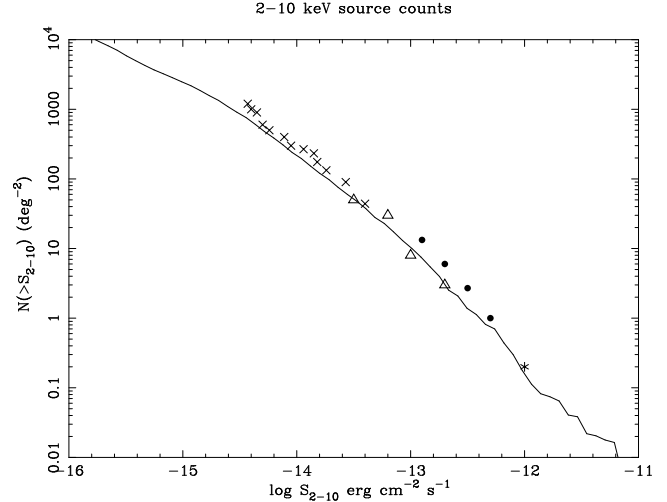}
\caption{\normalsize The solid line shows the source counts generated by the baseline model described in section~3, for comparison with the latest observations. The asterisk shows the result of the {\em Ginga} fluctuation analysis after removal of the cluster contribution (Butcher et al.~1997); the filled circles and open triangles are from the {\em ASCA} surveys of Ueda et al.~(1999) and Boyle et al.~(1998), respectively; the crosses are from the {\em Chandra} SSA13 field (Mushotzky et al.~2000).}
\label{fig:counts1}
\end{figure}

In Fig.~\ref{fig:counts1} we compare the 2--10\keV~source counts generated by our model with those observed. These include some {\em Chandra} results from Mushotzky et al.~(2000), which probe down to almost $10^{-15}$\ergpcmsqps~and essentially resolve the background flux in this band. 

\begin{figure}
\psfig{figure=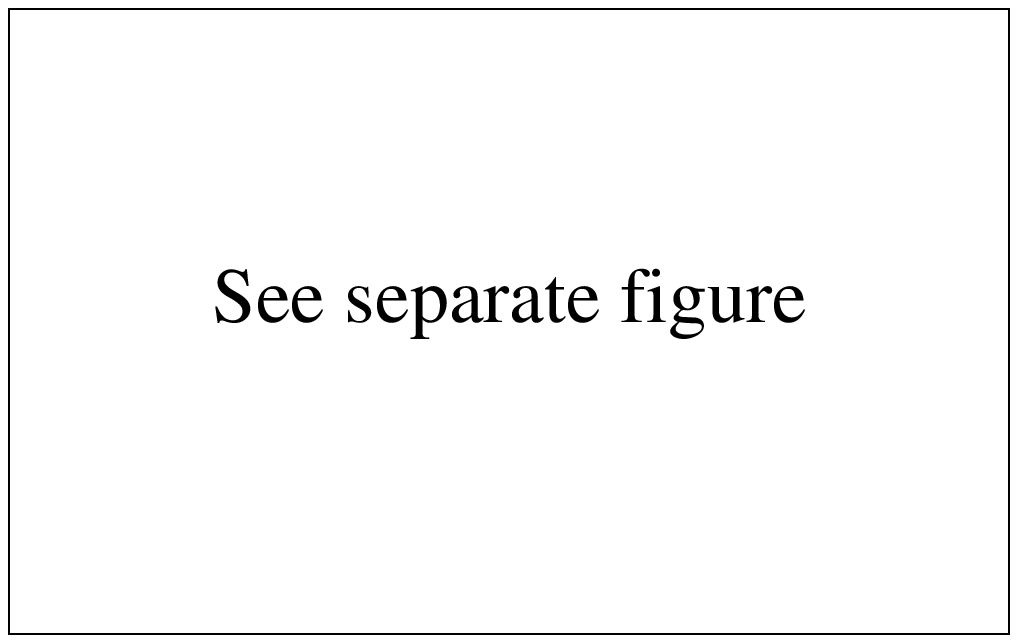,width=0.48\textwidth,angle=270}
\caption{\normalsize Plots of black hole mass versus host spheroid mass for all the AGN in the simulation: (a) immediately prior to the onset of the obscured accretion; (b) at the end of the obscured phase. The straight line is the locus $M_{\rm{BH}}=0.005M_{\rm{spheroid}}$ at which we halt the accretion if this has not already occurred by another means.}
\label{fig:MagAGN}
\end{figure}

As a further constraint, we calculate some limits on the contribution of our sources to the far-infrared and sub-millimetre backgrounds. The obscured objects alone make a (bolometric) contribution of 2.3 nW~m$^{-2}$~sr$^{-1}$, a figure which rises to 3.8 nW~m$^{-2}$~sr$^{-1}$ if the unobscured quasars are also included (but the SEDs of the latter objects peak in the optical/UV, not the far-infrared). These figures compare favourably with other XRB-based estimates of the AGN contribution to the far-infrared and sub-millimetre backgrounds made by Almaini et al.~(1999) and Fabian \& Iwasawa~(1999), and with the empirical constraints of McMahon et al.~(1999).

\section{DISCUSSION}
Having demonstrated that the model agrees satisfactorily with the observational data, we look more closely at its internal workings. A feature which distinguishes it from other semi-analytic models is that mergers between normal galaxies are not considered. To gauge how serious an omission this is, we list in Table 1 the fraction of haloes collapsing at different epochs which contain more than one normal galaxy and in which our model permits no further black hole growth. In computing these averages, note that we exclude systems from which supernovae have expelled the remaining gas (i.e. mainly dwarf galaxies where the black hole masses are just a few times the seed mass). Whilst the fraction of haloes with multiple normal galaxies is relatively small, the table also shows that such haloes contain, on average, significantly more massive black holes than isolated systems. This is because the black holes in the former haloes have in general already undergone one accretion episode taking them close to the Magorrian threshold (where $M_{\rm{BH}}=0.005M_{\rm{sph}}$), whilst most of the black holes in isolated systems are in newly-formed normal galaxies with black hole masses up to a factor of 5 below the Magorrian threshold (as discussed later in this section and illustrated in Fig.~\ref{fig:MagAGN}). If considered, merger-driven accretion power -- in which hot gas fuelling is supplemented by tidally-driven flows of cold gas towards the galactic nuclei -- would thus become increasingly significant towards the present epoch. As the most massive black holes get incorporated into groups/clusters, mergers could potentially supply the highest luminosity systems ($L_{\rm{X}} \geq 10^{45}$\ergps), of which there is a deficit when compared with observations (see Fig.~\ref{fig:xlf1}).

In order to understand the different evolutionary behaviour of the high and low luminosity systems in Fig.~\ref{fig:evol1}, Table 2 provides some statistics on the black hole masses and accretion rates (in Eddington units) at various epochs, and on the relative importance of the different termination mechanisms for the obscured growth phase. It can be seen that, at all epochs, most black holes have masses close to $10^{8}$\Msun~when the obscured Bondi accretion begins. Note that there is no accretion on to holes with $M_{\rm{h}}<2.6 \times 10^{7}$\Msun -- at least 16 seed holes (or, equivalently, 16 of the smallest mass haloes) must merge before any accretion can begin. At earlier times black holes are mainly in low mass dwarf galaxies, which by definition lack a cooling flow atmosphere and from which supernovae have often expelled the remaining gas. An epoch-dependent black hole mass therefore does not account for the evolution in Fig.~\ref{fig:evol1}. 

Fig.~\ref{fig:MagAGN} shows Magorrian diagrams -- plots of black hole versus spheroid mass -- for the AGN in the simulation, at the beginning and the end of the obscured phase, demonstrating that black holes grow in mass by up to a factor of 5 during this phase. With reference to eqn.~(3) for $N_{\rm{H}}$, we see that $N_{\rm{H}} \sim 3-15 N_{\rm{T}}$ at the start of the obscured phase, diminishing to $N_{\rm{H}} \sim 3 N_{\rm{T}}$ by its end. At every epoch, the majority of the obscured objects have $N_{\rm{H}} \sim 3 N_{\rm{T}}$. The vertical banding of Fig.~\ref{fig:MagAGN}(a) reflects the discrete nature of the block model and could be mitigated by using seed black holes with a range of mass (e.g. drawn from a gaussian distribution). Decreasing (increasing) the seed hole mass results in the accreting hole reaching the expulsion threshold, $M_{\rm{BH}}=0.005M_{\rm{spheroid}}$, at a later (earlier) epoch, leading to a higher (lower) XRB. If the seed mass is made too large, all holes are above the expulsion threshold {\em at formation} and there is no accretion.

In contrast, the accretion rate is strongly epoch-dependent: at $z<2.5$ most black holes are fed at less than 10 per cent of the Eddington rate, and at $z \leq 0.5$ a significant proportion is fed at less than 1 per cent of the Eddington rate; at $z \geq 2.5$, however, most holes accrete at between 10 and 50 per cent of the Eddington rate. As pointed out by NF00, the physical mechanism for this evolution is apparent from equation (8) of NF00 for the black hole growth rate; the latter is in general maximised for the hottest collapses, and since the virial temperature of a halo of mass $M$ collapsing at time $t_{\rm{coll}}$ scales as $(M/t_{\rm{coll}})^{2/3}$, it follows that, for a given mass, the earliest collapses give the most growth. We conclude that the luminosity-dependent evolution of Fig.~\ref{fig:evol1} is chiefly a reflection of an accretion rate which decreases strongly towards the present epoch, whilst the mass distribution of the accreting black holes depends far less strongly on redshift.

\begin{table*}
\begin{center}
\caption{The importance of normal galaxy mergers}
\begin{tabular}{llllll}
                     & $z<0.5$ & $0.5 \leq z < 1.5$ & $1.5 \leq z < 2.5$ & $2.5 \leq z < 4$ & $z \geq 4$ \\ 
$F(NG>1)^{\star}$  & 0.11    & 0.15               & 0.14             & 0.09             & 0.05       \\
${\overline M_{h(group)}^{\dagger}}$ ($10^{8}$ \Msun) & 5.2 & 5.5 & 5.3 & 4.8 & 3.7 \\
${\overline M_{h(isol)}^{\ddagger}}$ ($10^{8}$ \Msun) & 1.3 & 2.3 & 2.3 & 2.1 & 1.6 \\ 
\end{tabular}
\end{center}
$\star$ Fraction of collapsing haloes containing $>1$ normal galaxy. \\
$\dagger$ Mean black hole mass per normal galaxy in haloes collapsing with $>1$ normal galaxy. \\
$\ddagger$ Mean black hole mass in collapsing haloes with $\leq 1$ normal galaxy, prior to any accretion. \\
\end{table*}

\begin{table*}
\begin{center}
\caption{Statistics for accreting black holes}
\begin{tabular}{||llllll||} 
                     & $z<0.5$ & $0.5 \leq z < 1.5$ & $1.5 \leq z < 2.5$ & $2.5 \leq z < 4$ & $z \geq 4$ \\ 
\multicolumn{6}{c}{Fraction of obscured accretion phases ending by each means} \\ \\
Wind expulsion     & 0.88 & 0.96 & 0.98 & 0.98 & 0.93 \\
Hot gas exhaustion  & 0.02 & 0.03 & 0.003  & 0.0001 & 0.04 \\
New collapse      & 0.10 & 0.01 & 0.02 & 0.02 & 0.03 \\ \\ \\
\multicolumn{6}{c}{Mean black hole mass at start of the obscured growth phase ($10^{8}$\Msun)} \\ \\
                                       & 1.09          & 1.23           & 1.23          & 1.22          & 0.93          \\  \\ \\
\multicolumn{6}{c}{Distributions of black hole mass and accretion rate at start (end) of obscured growth phase$^{\dagger}$} \\ \\
$M_{\rm{h}}<10^{7}$\Msun               & 0 (0)         & 0 (0)          & 0 (0)         & 0 (0)         & 0 (0) \\
$10^{7} \leq M_{\rm{h}} < 10^{8}$\Msun & 0.35 (0.35)   & 0.11 (0.11)    & 0 (0)         & 0.001 (0)         & 0.29 (0) \\
$10^{8} \leq M_{\rm{h}} < 10^{9}$\Msun & 0.65 (0.65)   & 0.89 (0.89)    & 0.997 (0.997) & 0.995 (0.996) & 0.71 (0.998) \\
$M_{\rm{h}} > 10^{9}$\Msun             & 0 (0)         & 0.001 (0.001)  & 0.003 (0.003) & 0.004 (0.004) & 0.002 (0.002) \\ \\
$f_{\rm{Edd}} < 0.01$                  & 0.22 (0.42)   & 0.0007 (0.055)  & 0 (0.0002)     & 0 (0)  & 0 (0) \\
$0.01 \leq f_{\rm{Edd}} < 0.1$         & 0.78 (0.58)  & 0.97 (0.92)    & 0.62 (0.80)      & 0.065 (0.59) & 0 (0.22) \\
$0.1 \leq f_{\rm{Edd}} < 0.5$          & 0.0002 (0.0002) & 0.03 (0.024)    & 0.38 (0.195) & 0.932 (0.41) & 0.982 (0.77) \\
$f_{\rm{Edd}} > 0.5$                   & 0 (0)         & 0 (0)          & 0.0002 (0.0002) & 0.003 (0.003) & 0.018 (0.01) \\ 
\end{tabular}
\end{center}
$\dagger$ Each column shows the fraction of black holes in various intervals of mass and accretion rate (as a fraction $f_{\em{Edd}}$ of the Eddington accretion rate, $\dot{M}_{\rm{Edd}}= 4 \pi G m_{p} M_{\rm{h}} / c \epsilon \sigma_{\rm{T}}$), at the start of the obscured accretion phase and (in parentheses) at its end.
\end{table*}

Although Table 2 shows that wind expulsion is the primary mechanism for terminating the obscured growth of the black holes, the scheme may require some modification from the simple form envisaged by Fabian~(1999). He assumed that gas is ejected in a spherically symmetric manner, completely shutting off the black hole's fuel supply and enabling it to shine as an optically-selected quasar for no more than $\sim 10^{6}$ years while the accretion disk empties. The lifetime of the unobscured quasar in our model is, however, $\Delta t_{\rm{qso}}=9 \times 10^{7}$\yr, and comparable to the length of the optically bright quasar phase required by other modellers (e.g. Kauffmann \& Haehnelt~2000, Cattaneo~2000, Granato et al.~1999). If all the gas had been ejected, such a long accretion phase would be difficult to sustain unless the ejection were somehow anisotropic: ejection might for example occur along the hole's rotation axis, leaving a torus of centrifugally-supported, accretable material in the plane of the accretion disk. Such a scenario may simultaneously address the problem of the excess soft XRB produced by the model, as alluded to in section 3. This torus will plausibly lead to moderate absorption in lines of sight passing through it, as in orientation-based AGN unification schemes (see Antonucci~1993 for a review). The number of sources with absorbed X-ray spectra and narrow optical emission lines, compared to those appearing as classical quasars will thus depend upon the solid angle subtended by the outflow (or, equivalently, on the ratio of the times spent in the ejection phase and quasar phase proper, in those cases where the gas expulsion is isotropic and ultimately complete). Note that even anisotropic ejection is likely to clear the spheroid of all hydrostatically-supported gas, leaving only that which is centrifugally supported. This means that the overall picture in which ejection ultimately terminates both black hole and spheroid growth is retained.

We can also compute the mechanical energy deposited by the AGN winds in the surrounding gas during the obscured growth phase. Averaging over all objects in the simulation and taking a wind power of the form $L_{\rm{W}}=aL_{\rm{Edd}}$ (where $a$ is constant), we find that the winds deposit a mean specific energy of $1.4 \times 10^{51}a$ \erg \Msun$^{-1}$~into the surrounding gas of {\em their own} haloes (assuming that all the wind power is absorbed). If instead the energy is distributed evenly among all the baryons in the Universe, the specific energy input is slightly lower at $2.3 \times 10^{50}a$\erg \Msun$^{-1}$. For $a \sim 0.1$ (as suggested by Fabian~1999) these values are equivalent to $\sim 40$ and $\sim 7$\keV~per particle (for fully ionized gas with cosmic abundances), respectively, and are thus more than capable of providing the excess specific energy of 1.8--3.0\keV~per particle which Wu, Fabian \& Nulsen~(2000) find is required to break the self similarity of X-ray cluster properties and match the observed $L_{\rm{X}}-T$ relation. That AGN winds may be important in this regard has been recognised by Ensslin et al.~(1998), Wu et al.~(2000) and Bower et al.~(2000). A discussion of the impact of this energy release on the general (i.e. non-cluster) intergalactic medium is deferred to a later paper.

\section{CONCLUSIONS}
We have developed a physical model for the hard XRB which works {\em forward} in cosmic time from the epoch when the first massive bound objects were forming, by bringing together a semi-analytic galaxy formation model and prescriptions for the fuelling and obscuration of massive black holes. Quasars are assumed to be fuelled by hot gas from the cooling flow atmospheres which form around collapsing objects about the size of normal galaxies, and accretion is terminated by wind-driven gas expulsion when the mass of a black hole reaches 0.005 times that of its host spheroid (thereby forcing agreement with the findings of Magorrian et al.~1998). The model accounts quantitatively for the spectrum of the XRB and the source counts in the 2--10\keV~band, and is compliant with current constraints on the AGN contribution to the far-IR and sub-mm backgrounds. Qualitatively at least, it also reproduces the dramatic decline in the co-moving density of luminous quasars since $z=2$. Our aim was merely to demonstrate that, with simple assumptions and reasonable values for the free parameters, many observations can be satisfactorily reproduced. A number of deficiencies in the model remain, such as our use of the relatively crude block model for hierarchical structure formation, and in the astrophysical details of the fuelling and obscuration via the cooling flow and its cold clouds (as mentioned in the final paragraph of section 2.2).

With its emphasis on {\em hot gas} fuelling and neglect of mergers between normal galaxies, our formalism may be considered orthogonal to that adopted by Cattaneo~(2000) and Kauffmann \& Haehnelt~(2000), where quasars are fuelled by cold gas in major mergers with the host spheroid being formed in the process. It would clearly be of interest to construct a model which utilises elements from both approaches.

\section*{ACKNOWLEDGMENTS} RJW acknowledges a PPARC Studentship and thanks Kelvin Wu for assistance during the early stages of this project. ACF thanks the Royal Society for support. We thank the referee, Gianni Zamorani, for a careful reading of the manuscript and constructive comments.

{}

\end{document}